\documentstyle[astrobib,psfig]{mn-ab}

\def\rvir{R_{\rm vir}}
\def\rt{R_{\rm t}}
\def\tz{\tilde z}
\def\omm{\Omega_{\rm m}}
\def\oml{\Omega_{\Lambda}}
\def\Dvir{\Delta_{\rm vir}}
\def\lcdm{$\Lambda$CDM}
\def\hmpc{\ h^{-1}{\rm Mpc}}
\def\rs{R_{\rm s}}

\def\Msun{M_{\odot}}
\def\hMsun{h^{-1} \Msun}
\def\hsolmass{h^{-1}M_{\odot}}
\newcommand{\nbody}{$N$-body~}

\title[Origin and Destiny] 
{Origin and Destiny of Dark Matter Halos: \\  
    Cosmological Matter Exchange and Metal Enrichment}
\author[Kolatt et al.]
{T.S. Kolatt$^{1,2}$, 
J.S. Bullock$^{2,3}$, 
A. Dekel$^1$,
J.R. Primack$^2$, \cr
Y. Sigad$^1$, 
A.V. Kravtsov$^{3,4}$, 
\&
A.A. Klypin$^4$ \\
$^1$Racah Institute of Physics, The Hebrew University, Jerusalem, 
    91094, Israel\\
$^2$Physics Department, University of California, Santa Cruz, CA 95064 USA\\
$^3$Department of Astronomy, Ohio State University, Columbus, OH 43210 USA\\
$^4$Astronomy Department, New Mexico State University, Box 30001,
            Dept. 4500, Las Cruces, NM 88003-0001 USA} 

\begin{document}

\maketitle


\begin{abstract}

We analyze  the exchange of dark  matter between halos,  subhalos, and
their environments in a high-resolution cosmological $N$-body
simulation  of a  \lcdm\ cosmology.   
At each analyzed redshift $z$ we divide  the dark
matter particles into 4 components: 
(i) 
isolated 
galactic halos, (ii) subhalos, (iii) the diffuse medium of group and cluster
halos, and (iv) the background outside of virialized halos. 
We follow the time evolution of the mass distribution and flows
between these components and provide fitting functions for the exchange
rates.

The exchange rates show gradual evolution 
as $z$ decreases to 2, and become more
steady thereafter. 
For $z \la 2$ about $15\%$ of the isolated galactic halos cluster per
Gyr to become subhalos and a similar fraction of their mass returns to the 
unvirialized background. 
Mass accumulation onto subhalos is equally shared between previously
isolated halos and unvirialized matter, and is dominated by
accretion from the host's diffuse matter beyond $z\simeq1$.
This accumulation is balanced for $z\simeq0.5$ by subhalo disruption 
at a rate of about half of their mass per Gyr. 
The diffuse component in host halos
is built by accreting isolated halos and un-virialized material in
mass shares of 40\% and 60\%, respectively, 
and at $z<0.5$ also by disruption of
subhalos.
The unvirialized IGM is enriched mostly by stripping of
isolated halos, and at $z<1$ also by mass loss from 
groups and clusters.

We go on to use our derived exchange rates together with a simple recipe for
metal production to gauge the importance of metal redistribution in
the universe due solely to gravity-induced
interactions.
This crude model predicts some trends regarding metallicity ratios. 
The diffuse metallicity in
clusters 
is predicted to be $\sim 40\%$ that in isolated
galaxies ($\sim 55\%$ of groups) at $z=0$, and should be lower 
only slightly by $z=1$, consistent with observations.  
The metallicity of the
diffuse media in large galaxy halos and poor groups
is expected to be lower by about a factor of $5$ 
by $z\sim 2$, in agreement with the
observed metallicity of damped Ly$\alpha$ systems.  The metallicity of
the background IGM is predicted to be $(1-3)\times10^{-4}$ that of
$z=0$ clusters, also consistent with observations. The agreement of 
predicted and observed trends indicates that 
gravitational interaction alone
may play an important role in metal 
enrichment of the intra-cluster and intergalactic media.

\end{abstract}

\begin{keywords}
galaxies: formation -- galaxies: evolution -- cosmology: theory --
cosmology:dark matter
\end{keywords}

\section{Introduction}
\label{sec:intro}


Under the standard assumption of 
hierarchical galaxy formation,  
galaxies reside within virialized dark-matter (DM) halos.  
In this picture, field galaxies can be identified with ``isolated''
galaxy-size halos, 
while cluster, group galaxies, and even satellites of massive galaxies,
can be identified with ``subhalos'' 
embedded in the background of a larger ($\ga 10^{13} \hMsun$) 
``host'' halo.  

We can make a further association by identifying
the intra-cluster medium with the diffuse
matter of massive ``host halos'' --- the mass between the subhalos,
and the low-density  intergalactic medium (IGM) with
matter located outside of any (galactic-sized
or bigger)
halo~\footnote{
Although the IGM is generally associated with a gas component,
we are making the assumption that this diffuse gas is
tracing the diffuse dark matter.}.

Several phenomena suggest that matter is continuously exchanged between 
grouped galaxies, field galaxies, and different regions of the IGM.
These include cooling flows in clusters, groups and
elliptical galaxies \cite{fabian_rev:94,fabian_nulsen:94}, 
and galaxy interactions \cite{barnes_rev:92}.
This interchange of matter between different populations is
qualitatively  expected within the CDM scenario,
where halos are continuously interacting and merging into larger halos.
However, the analytic descriptions of this process, such as the Press-Schechter
formalism (PS) and its variants \cite{press:1974,lc:93,lc:94},
fail to match the
observed phenomena in two important ways: 
(1) they do not follow the halos as distinct
entities once they are incorporated in larger halos,
and (2) they treat accretion into halos as a one-way process, ignoring
expulsion back into the diffuse media due to heating and tidal forces.
Evidence for the
existence of such phenomena and the 
survival of subhalos is provided, for example,
by cluster galaxies' velocity profiles (e.g., \citeNP{amram:94}) and 
halo truncation observed
via high-resolution 
density reconstructions by gravitational lensing in clusters
\cite{natarajan:98,tyson:98}. Another indication for expulsion 
of matter from subhalos
comes from the relatively high metallicity observed in the 
hot diffuse matter in clusters.  

One way to improve analytic treatments is by semi-analytic modeling, 
in which complex processes 
(such as gas cooling, star formation, and supernova feedback) 
in galaxies within merging dark matter halos
are followed via simplified recipes.
As argued below, there are several limitations to current 
semi-analytic models. 
In particular, the nonlinear substructure 
can be properly resolved only via full-scale cosmological simulations,
which also provide the spatial information missing in 
semi-analytic models.

The usual semi-analytic approach (e.g., \citeNP{kwg:93,cafnz:94,sp:99})
using extended PS merging trees does not take into account mass loss 
and exchange between interacting halos.
Certain aspects of galaxy formation may be severely affected by
these missing features.  For example, different accretion rates
onto clustered galaxies and field galaxies should affect their
relative star formation rates.  In addition, 
the transfer of material that has been ``processed'' in galaxies
into the diffuse media may act to transfer metals and heat
into the unvirialized IGM (see below.)

Only recently has the dynamic range of \nbody\ simulations become wide
enough to allow 
resolution of
halo sub-structure in a cosmological statistical
sample (e.g., \citeNP{klypin:overcoming}).
By utilizing such \nbody\ experiments it is possible to investigate
the properties of the hierarchy of halos and diffuse matter
\cite{kolatt-lbg,bullock_thesis,profiles}.
For example, we present elsewhere (\citeNP{collision_rate})
an analysis of collision rates of sub-structure.

In this paper, we study the exchange of matter
using such a simulation.
We quantify the flow of dark matter among four
components: isolated halos,  subhalos, the diffuse media of 
host halos, and the  unvirialized background.
These categories 
roughly correspond to field galaxies, grouped galaxies, the intra-cluster 
medium, and the background IGM.
The aim is to add a solid 
quantitative result to the general expectation of matter
exchange in the hierarchical picture, and also
to identify the important 
exchange processes which might provide input for future
modeling.

An important example of a process where our results are relevant
is the
large-scale redistribution of metals in the  universe caused solely by
gravitational interactions. 
In the second half of the paper we demonstrate
how our derived exchange rates 
shed light on  this issue using a
crude but explicit model for 
metal enrichment.  Our model
assumes that supernova winds
spread 
processed 
galactic gas 
from the disk 
uniformly throughout galactic halos (galaxy-mass
subhalos and halos).  We associate a fraction of each  unit of dark
mass with gas and assign to it a metallicity in proportion to the time
it spends  inside a  galactic halo.   By  following the  flow  of this
``enriched gas'', we estimate  the effect of gravitational exchange on
enriching the  IGM  and the  diffuse media  of   clusters, groups, and
massive galaxy halos.  We make predictions  for the relative abundance
of these populations and study how the metallicities should evolve with
redshift.

In \S\ref{sec:simu} we provide a brief description of the simulation, 
the halo finder, and the construction of the halo hierarchy.
In \S\ref{sec:rates} we quantify statistically the matter exchange rates
between the four components
and discuss the origin and destiny of the matter in each component.
In \S\ref{sec:metal} we address the evolution of metallicity in the different
components and compare to observational measurements. 
We discuss our results and conclude in \S\ref{sec:conc}.

\section{Simulated Halos}
\label{sec:simu}

We used the ART code \cite{kkk:97} to simulate the evolution of 
collisionless DM in the ``standard" \lcdm\ model 
($\Omega_{\rm m}=1-\Omega_{\Lambda}=0.3$; $H_0=100h=70$ 
km s$^{-1}$ Mpc$^{-1}$; 
$\sigma_8=1.0$).
This model universe has a present age $t_0 = 13.5$ Gyr.
The simulation followed the trajectories of $256^3$ particles
within a cosmological periodic box of size $L = 60 \hmpc$  
from redshift $z = 40$ to the present.  
A basic $512^3$ uniform grid was used, and six
refinement levels were introduced in the regions of highest density, implying
a dynamic range of $\sim 32,000$.  The formal resolution of the simulation
is thus $f_{\rm res} \approx 2 {\rm h^{-1}kpc}$, and the mass per
DM particle is $m_{p} \approx 1 \times 10^{9} \hMsun$.
We analyze 15 saved outputs at times between $z=5$ and $z=0$.

The identification of halos is a key feature of the analysis;
we try to make it objective and self-consistent, 
following the evolution involving halo interactions and mergers.
Traditional halo finders utilize either friends-of-friends algorithms or
overdensities in spheres or ellipsoids to identify virialized
halos. These algorithms fail to identify sub-structure
with well-defined attributes and errors. 
We therefore have designed a new hierarchical halo finder, based on the bound
density maxima (BDM) algorithm \cite{klypin:overcoming}. 
The details of the halo finder are described elsewhere 
\cite{bullock_thesis,profiles}, and we summarize below only its main 
relevant features.

After finding all 
density maxima in the simulation,
we unify overlapping maxima, define a minimum number of 
particles per halo ($N_{\rm p}^{\rm min}$),
and iteratively find the center of mass of a sphere about 
each of the remaining maxima.
We compute the spherical density profile about each center 
and identify the halo virial radius $R_{\rm vir}$
inside which the mean overdensity has dropped to a value $\Dvir$,
based on the spherical infall model. 
For the family of flat cosmologies ($\omm+\oml=1$),
the value of $\Dvir$ can be approximated by (\citeNP{bryan:98})
$\Dvir \simeq (18\pi^2 + 82x - 39x^2)/(1+x)$,
where $x\equiv \Omega(z)-1$.
In the \lcdm\ model used in the current paper, $\Dvir$ varies from
about 180 at $z\gg 1$ to $\Dvir\simeq 340$ at $z=0$.
If an upturn occurs in the density profile inside $\rvir$,
we define there a truncation radius $\rt$.

An important step of our procedure is the fit of the density profile 
out to the radius $\min(\rvir ,\,\rt)$ with a universal functional form.
We adopt the NFW profile \cite{nfw:95},
\begin{equation}
\rho_{\rm{NFW}}(r) = \frac{\rho_s}{(r/\rs)\left(1+r/\rs\right)^2},
\label{eq:nfw}
\end{equation}
with the two free parameters $\rs$ and $\rho_s$ --- a characteristic
scale radius and a characteristic density. This pair of parameters could
be equivalently replaced by other pairs, such as $\rs$ and $\rvir$.
Using this fit, we iteratively remove unbound particles from each modeled 
halo and unify every two halos that overlap in their $\rs$ and are 
gravitationally bound.
Finally, we look for virialized regions within $\rs$ of big halos to
identify subhalos near the centers of big host halos, e.g., mimicking
cD galaxies in clusters.
The minimum halo mass corresponding to 
$N_{\rm p}^{\rm min}=50$ 
particles is $\sim 5\times 10^{10}\hsolmass$. 
At this minimum mass though, the finder is incomplete and fails to
identify some halos. Completeness (i.e., $100\%$ halo identification) is
reached at $M \sim 1.5 \times 10^{11} \hsolmass$ (cf. Sigad et al.
2000).
The modeling of the halos with a given functional form allows us
to assign to them characteristics such as a virial mass and radius,
and to estimate sensible errors for these quantities.

The classification scheme used in the rest of the paper is as follows.
A {\it subhalo} is a halo 
whose its center 
lies within the virial radius
of a larger halo~\footnote{``Larger'' means the host is
at least 25\% more massive than the subhalo. In the current 
application we limit ourselves to the first level of
subhalos, but the classification scheme can straightforwardly be
extended to deal with many levels of subhalos within subhalos.}.
A {\it host} is a halo that contains at least one subhalo.
An {\it isolated} halo is any halo that is not 
a subhalo and is also not a host.  Finally, in
\S~\ref{sec:metal}, the combined 
set 
of subhalos and
isolated halos are identified as {\it galactic} halos.  All
of the mass that is not contained within any identified halo
is referred to as {\it unvirialized}.
Note that in the PS formalism subhalos are not taken to be independent
quantities, and that without a lower mass cutoff for halos, the 
unvirialized component is not well defined.

\section{Matter exchange rates}
\label{sec:rates}

At any given output time, we assign each mass particle to one of four 
components, and label each of them as follows: 
\begin{enumerate}
\item
Isolated halos (I)
\item
Subhalos (S)
\item
Diffuse matter in host halos (D)
\item
``Unvirialized'' diffuse matter (U)
\end{enumerate}
In the next section we will relate the first two components 
to galaxies 
and the last two (diffuse components) to intergalactic gas, 
but in this section we simply study the exchange rates of mass among
the four components.

The division into components clearly depends on the details of the halo finder, 
and in particular on the minimum mass imposed. For example, had this mass 
been set to be as small as the particle mass, all the particles would have been
associated with ``virialized halos.'' 
Possible incompleteness of our halo
finder near the minimum mass may slightly affect the subhalo population
but this is not a major concern here because 
the subhalo mass function is somewhat flatter than the distinct halo 
mass function \cite{sigad:00} such that most of the mass is in
subhalos more massive than the completeness limit of $10^{11}\hsolmass$.

At every output time, we compute 
the total mass in each component as well as the rates
of mass exchange between the components, all per unit volume.
We denote the average density corresponding to component $x$ by $\rho_x$, 
and the rate of density flow from component $x$ to component $y$ by 
$\dot \rho_{xy}$. The net flow between the two components is thus
$\dot \rho_{xy} - \dot \rho_{yx}$.
The exchange rate is estimated by counting the amount of mass gained and 
lost between
each component and dividing by the elapsed 
proper 
time between outputs.
The timesteps are 
selected to be typically separated by $\sim$1 Gyr.  
In the following, the mean densities are referred to in comoving units of 
$10^{-2}\langle \rho \rangle$, where $\langle\rho \rangle=8.3\times10^{10}
\hMsun (\hmpc)^{-3}$ is the mean comoving density of the
simulated cosmological model.
The flows are referred to correspondingly in units of
$10^{-2}\langle \rho \rangle {\rm Gyr}^{-1}$.

\begin{figure*}
\vskip-3truecm
\centerline{\psfig{file=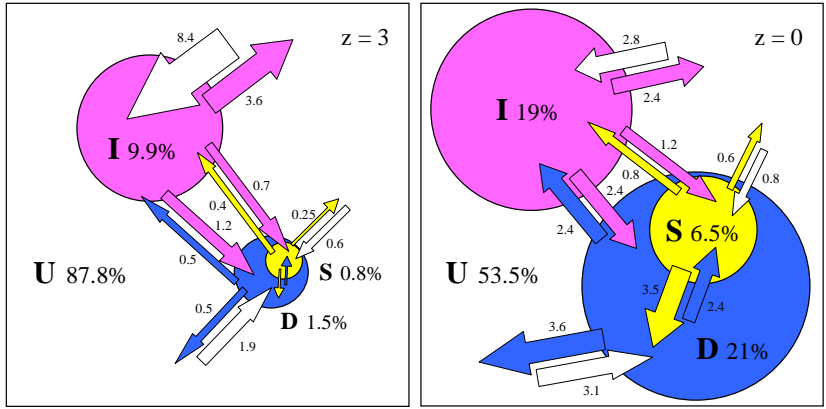
 ,height=18.15truecm,width=14truecm }}
\vskip-9truecm
\caption{
Mass distribution among the four components and exchange rates.
U=unvirialized background, I=isolated halos, D=diffuse in host halos,
S=subhalos. 
Mean densities are quoted in percentiles (i.e., units of $10^{-2}\langle
\rho \rangle$),
and flows in $10^{-2}\langle \rho \rangle$ Gyr$^{-1}$.
The areas are proportional to the mass in the corresponding component.
The thickness of the arrows is monotonic with the flow.
Left: $z=3$. Right: $z=0$.
}
\label{fig:pop}
\end{figure*}

\begin{figure}
\centerline{\psfig{file=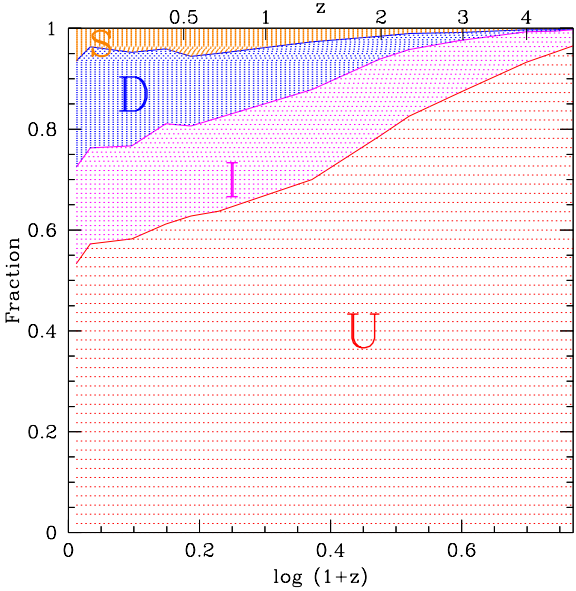
 ,height=8truecm,width=8truecm }}
\caption{
Fractional density of each one of the four components U,I,D,S (cf. Fig.
\ref{fig:pop}) as function of redshift.
}
\label{fig:den_evolution}
\end{figure}

Figure~\ref{fig:pop} is a schematic diagram showing 
the distribution of mass among the 4 components and 
the flows between them, at two different times, $z=0$ and 3. 
The area associated with each component is
proportional to the mass in that component, and the thickness of the
arrows is a monotonic function of the flow. 
Figure \ref{fig:den_evolution} depicts the redshift evolution of 
the fractional density in each one of the components.

\begin{figure}
\centerline{\psfig{file=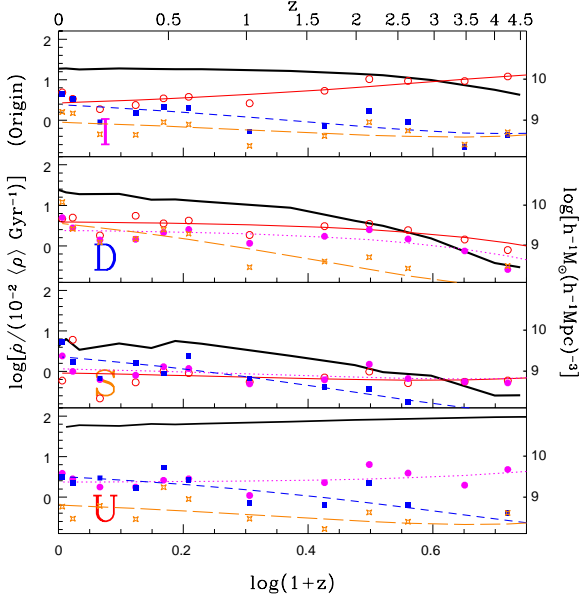
 ,height=8truecm,width=8truecm }}
\caption{
Time evolution of the mass density in each of the 4 components
(thick black curve in each panel) 
in units of $\hsolmass(\hmpc)^{-3}$ (right axis) or
equivalently $10^{-2}\times \langle \rho \rangle$, the comoving 
average density of the simulated cosmology (left axis, dropping the
``Gyr$^{-1}$''). 
Also plotted are the incoming flow rate into each component from 
each of the other 
components (``Origin'') in units of $10^{-2} \times \langle \rho \rangle
$Gyr$^{-1}$ (left axis).
The symbols are the measured flow in the given time interval,  
and the curves are quadratic fits.
Isolated halos I=filled circles, dotted lines, magenta;
diffuse matter in host halos D=filled squares, short-dashed lines, blue;
subhalos S=empty stars, long-dashed lines, orange;
unvirialized matter U=open squares, solid lines, red.
}
\label{fig:origin}
\end{figure}

\begin{figure}
\centerline{\psfig{file=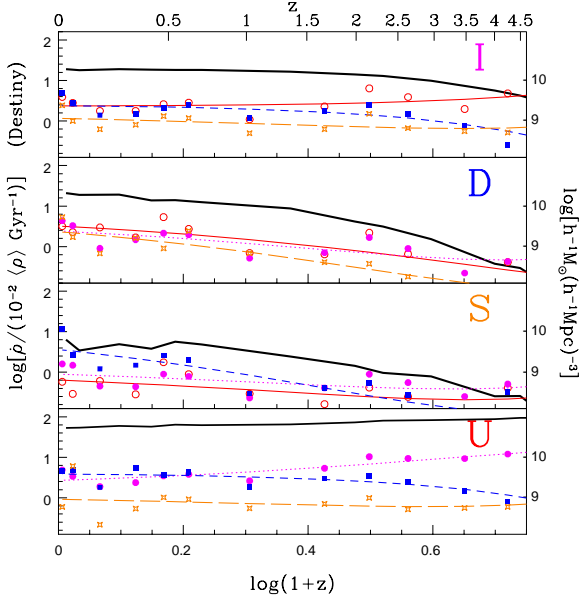
 ,height=8truecm,width=8truecm }}
\caption{
Destiny: same as Fig. \ref{fig:origin}, but showing the
outgoing flow rate from each component into each of the other components.
}
\label{fig:destiny}
\end{figure}

In presenting the exchange rates
we can either focus on the {\it origin} of the incoming mass to each 
component or on the {\it destiny} of the outgoing mass from each component.
The information content is redundant, because the $x$  origin of $y$ is
the same as the $y$ destiny of $x$
in the previous time-step, 
but they allow two different angles of view.
Figures \ref{fig:origin} and \ref{fig:destiny} summarize the exchange 
rates in these two ways.
They show the evolution of total mass density in each of the four 
components, and the flow rate into (from) each component from (into)
each of the other components.  We note that 
the general variation in time of many of the rates is slow.
We fit this weak 
time dependence of the measured rates with a quadratic function in 
$\tz \equiv \log (1+z)$,
\begin{equation}
\log \dot\rho = \log(\dot\rho_0)  + b \tz + a \tz^2 \,.
\end{equation}


\begin{tabular}{|c|c|c|c|}
\multicolumn{4}{c}{Table 1} \\
\hline
\multicolumn{4}{|c|}{Density fit coefficients ($\log{\rho_0}$),b,a$^{*}$} \\
\hline\hline
\multicolumn{1}{|c|}{ I } 
&\multicolumn{1}{c|}{ D } 
&\multicolumn{1}{c|}{ S } 
&\multicolumn{1}{c|}{ U } \\ \hline
\multicolumn{1}{|c|}{10.132 } 
&\multicolumn{1}{c|}{10.154} 
&\multicolumn{1}{c|}{9.591} 
&\multicolumn{1}{c|}{10.439} \\
\multicolumn{1}{|c|}{0.906} 
&\multicolumn{1}{c|}{0.656} 
&\multicolumn{1}{c|}{0.497} 
&\multicolumn{1}{c|}{0.092} \\ 
\multicolumn{1}{|c|}{-2.287}      
&\multicolumn{1}{c|}{-4.218}     
&\multicolumn{1}{c|}{-3.220}     
&\multicolumn{1}{c|}{-0.013} \\ \hline
\multicolumn{4}{l}{$\tilde z=\log(1+z)$} \\
\multicolumn{4}{l}{$^{*}\log({\rho}_{x})=\log(\rho_0)+ b\tilde z+
a\tilde z^2$ } \\
\multicolumn{4}{l}{[${\rho}$]=h$^{-1}M_\odot$(h$^{-1}$Mpc)$^{-3}$}  \\
\end{tabular}


\vskip0.5truecm

We assign equal weights to the measured rates in the different time steps; 
this is under the assumption that the errors are dominated by cosmic variance, 
which we are not trying to model in detail. 
Table 1 displays the parameters for the quadratic functional fits to 
the density evolution of each component, $\rho(\tz)$, the fits
are accurate to $10-15\%$, and
Table 2 displays the parameters for the functional fits to 
the whole matrix of flows, $\dot \rho_{xy}$. 
For example, from I to D, $\log \dot\rho_{ID}(\tz) = 9.306 + 0.043 \tz 
- 1.242 \tz^2$, which equals 8.88 at $z=3$; thus $\dot\rho_{ID} = 
2.02 \times 10^9$ at $z=0$ and $7.6 \times 10^8$ at $z=3$, in units of 
$h^{-1} M_\odot (h^{-1} {\rm Mpc})^{-3}$ Gyr$^{-1}$.
In general, the evolution slows down with time because of the 
\lcdm\ cosmology; with $\Omega_{\rm m}=0.3$, the characteristic 
epoch for loitering is $z \sim 0.7$.


\begin{tabular}{|c c|c|c|c|c|}
\multicolumn{6}{c}{Table 2} \\
\hline
\multicolumn{6}{|c|}{Rate fit coefficients ($\log{\dot \rho_0}$),b,a$^{*}$} \\
\hline\hline
\multicolumn{2}{|c|}{
\put (10.0,4.5){\small{To$\rightarrow$}}
\put (1.0,-5.0){\small{From}}
}
&\multicolumn{1}{c|}{ I } 
&\multicolumn{1}{c|}{ D } 
&\multicolumn{1}{c|}{ S } 
&\multicolumn{1}{c|}{ U } \\ \hline
\multicolumn{2}{|c|}{}
&\multicolumn{1}{c|}{} 
&\multicolumn{1}{c|}{9.306} 
&\multicolumn{1}{c|}{9.027} 
&\multicolumn{1}{c|}{9.347} \\
\multicolumn{2}{|c|}{\hskip0.2truecm I \hskip0.2truecm }
&\multicolumn{1}{c|}{-} 
&\multicolumn{1}{c|}{0.043} 
&\multicolumn{1}{c|}{-0.872} 
&\multicolumn{1}{c|}{-0.615} \\ 
\multicolumn{2}{|c|}{}
&\multicolumn{1}{c|}{ }      
&\multicolumn{1}{c|}{-1.242}     
&\multicolumn{1}{c|}{0.694}     
&\multicolumn{1}{c|}{1.225} \\ \hline
\multicolumn{2}{|c|}{}
&\multicolumn{1}{c|}{9.367} 
&\multicolumn{1}{c|}{} 
&\multicolumn{1}{c|}{9.305} 
&\multicolumn{1}{c|}{9.408} \\ 
\multicolumn{2}{|c|}{\hskip0.2truecm D \hskip0.2truecm }
&\multicolumn{1}{c|}{-1.621} 
&\multicolumn{1}{c|}{-} 
&\multicolumn{1}{c|}{-1.673} 
&\multicolumn{1}{c|}{-0.621} \\ 
\multicolumn{2}{|c|}{}
&\multicolumn{1}{c|}{0.776}      
&\multicolumn{1}{c|}{}     
&\multicolumn{1}{c|}{-0.349}     
&\multicolumn{1}{c|}{-1.212} \\  \hline 
\multicolumn{2}{|c|}{}
&\multicolumn{1}{c|}{8.961} 
&\multicolumn{1}{c|}{9.580} 
&\multicolumn{1}{c|}{} 
&\multicolumn{1}{c|}{8.682} \\ 
\multicolumn{2}{|c|}{\hskip0.2truecm S \hskip0.2truecm }
&\multicolumn{1}{c|}{-1.589} 
&\multicolumn{1}{c|}{-2.965} 
&\multicolumn{1}{c|}{-} 
&\multicolumn{1}{c|}{-0.462} \\ 
\multicolumn{2}{|c|}{}
&\multicolumn{1}{c|}{1.436}      
&\multicolumn{1}{c|}{0.957}     
&\multicolumn{1}{c|}{}     
&\multicolumn{1}{c|}{-0.258} \\ \hline  
\multicolumn{2}{|c|}{}
&\multicolumn{1}{c|}{9.390} 
&\multicolumn{1}{c|}{9.492} 
&\multicolumn{1}{c|}{8.925} 
&\multicolumn{1}{c|}{} \\ 
\multicolumn{2}{|c|}{\hskip0.2truecm U \hskip0.2truecm }
&\multicolumn{1}{c|}{0.109} 
&\multicolumn{1}{c|}{0.233} 
&\multicolumn{1}{c|}{-0.842} 
&\multicolumn{1}{c|}{-} \\ 
\multicolumn{2}{|c|}{}
&\multicolumn{1}{c|}{1.086 }      
&\multicolumn{1}{c|}{-1.277}     
&\multicolumn{1}{c|}{0.830}     
&\multicolumn{1}{c|}{} \\  \hline 
\multicolumn{6}{l}{$\tilde z=\log(1+z)$} \\
\multicolumn{6}{l}{$^{*}\log({\dot \rho}_{xy})=\log(\dot \rho_0)+ b\tilde z+
a\tilde z^2$ } \\
\multicolumn{6}{l}{[${\dot \rho}$]=h$^{-1}M_\odot$(h$^{-1}$Mpc)$^{-3}$Gyr$^{-1}$ }  \\
\end{tabular}



\subsection{Total mass distributions}

%
The main features of the {\it total mass} evolution in each component 
in the redshift interval $z=4.5$ to 0 are as follows:
\begin{enumerate}
\item
The 
isolated component, I, 
grows from 5 to 19, with most of the relative 
growth occurring at $z>2$. 
\item
The diffuse component, D,  grows continuously from 0.2 to 19.
The growth continues to be significant down to $z\sim1$,
long past the time when I component growth stagnates.
\item
The subhalo component S grows at roughly a constant rate 
until $z\sim 1$, from 0.2 
to 3.6, and then becomes rather constant. 
\item
The background U is depleted slowly, from a density of 87 before $z=3$ 
to 53 at $z=0$. 
\end{enumerate}

\subsection{Origin and Destiny}

The main characteristics of the {\it origin} of each component are as follows:
\begin{enumerate}
\item
The origin of I (isolated halos) is dominated by accretion from 
U (unvirialized component) at a gradually 
decreasing rate, from 12 to 2.6 
(recall that the flow units are $10^{-2}\langle \rho \rangle$Gyr$^{-1}$).
About 100\% of I is being added every Gyr at $z\sim 3$, while
this fraction drops to $\sim 20\%$ at $z=0$.
(Note also that $\sim$10\% of I comes from D and S; this is mostly 
a consequence of small host halos turning into simple halos because of the 
disruption of all their subhalos).
\item
The main source of D (diffuse matter in host halos)
is accretion from U, at a slowly varying rate, 
from 1.1 to 3.6.
In parallel, there is an important supply to D from infalling I, 
which is always about 1/3 of the total incoming flow.
About 100\% of D is being added every Gyr at $z\sim 2.5$,
dropping to 40\% at $z=0$.
The expulsion rate from S into D, 
corresponding to the disruption of subhalos, grows continuously
after $z\sim 3$, and 
becomes higher than the infall rate from I into D
at $z<0.4$, and comparable to the accretion from U at $z=0$.
\item
The origin of S at $z>1$ is divided about equally between I and U,
at roughly a constant rate of $\sim 1$ each. 
At $z<1$, the origin of S becomes dominated by accretion from D, 
reaching a rate of 2.5 at $z=0$.
The input to S is about 100\% of S per Gyr at $z\sim 2.5$, and 70\% 
after $z=2$.
It is likely that 
fraction of the accretion to S from U
is due to halos that formed outside of a host and
subsequently fell in, however the time resolution of the
simulation outputs analyzed
did not allow the identification of the intermediate
I component.  
The same applies to the fraction of matter that went to S through a D
phase.
\item
The little input to U comes mostly from expulsion of I, 
and at $z<0.5$ also from D, at the level of a few percent of U per Gyr. 
\end{enumerate}

The main features of the {\it destiny} of each component are as follows:
\begin{enumerate}
\item
Isolated halos,  I 
expel mass mostly into U, at roughly a constant rate of 2.5-3.5.
I also turns into D at a growing rate that becomes comparable to 2.5 at $z<1$.
The latter 
is likely due to I halos falling into groups and clusters 
and then being disrupted into D within a single timestep, without
ever being identified as S subhalos.
At $z\sim 4.5$, about 100\% of I is being lost (to U) per Gyr,
while after $z=2$ the outgoing rate (to U and D) is about a 1/3 of I.
This is
as expected from the relative masses of the isolated halos at the two
redshifts \cite{sigad:00}, 
less concentrated 
density profiles at higher redshift \cite{profiles}, and a higher
merger rate before $z\simeq2$ \cite{kolatt-lbg}. 
\item
The diffuse component 
D outputs mass in roughly equal parts to U and I (the latter being
mostly D+S turning into I).
At $z<1$, a significant part of the outgoing mass goes also to S.
The output rate from D to U grows from 0.3 to 3.
The total output rate is about 250\% of D per Gyr at $z\sim 4.5$,
and about 1/3 of D after $z=2$.
\item
The disruption of subhalos causes S to lose roughly 
a constant $\sim$40\% of its total mass every Gyr to D,
with a rate growing from 0.1 to 3.            
There is also some outflow from S to U at roughly a constant rate of
0.36-0.6, and 
transition from S to I at a rate 0.7-1.0.
\item
The main destiny of U is I. The accretion from U to D becomes comparable
after $z=1$, at a total accretion rate into halos of 6, which is $\sim
10\%$ of U per Gyr.
\end{enumerate}

The largest absolute flow at $z>2$ is accretion from U to I, at a rate 
of $\sim 10$,
partly balanced by a flow back from I to U at a rate of $\sim 4$.
They reach a near balance by $z=0$, at a rate of $\sim 2.5$.
At $z<1$, the dominant flow is the accretion from U to D, at a rate of
$\sim 4$, which is approaching a balance with the back-flow from D to U only
near $z=0$.

At $z>2$, all the halo components (especially I and D) absorb mass at a 
relative rate comparable to their own mass every Gyr, and lose mass at 
a lower rate.
At $z<2$ the fractional mass inflow is typically 10 to 50\% of each
halo component per Gyr, and the outflow rate is only slightly lower.
The S component exchanges mass at a high rate of 40-70\% of its own mass
every Gyr.

\subsection{Net mass exchange rates}

\begin{figure}
\centerline{\psfig{file=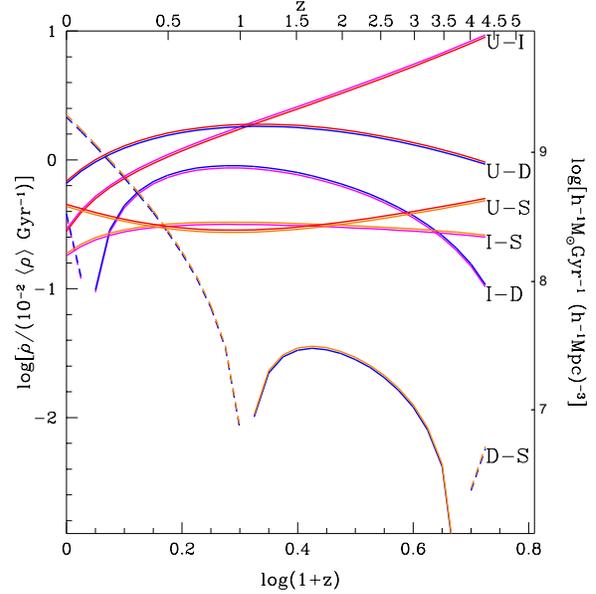
 ,height=8truecm,width=8truecm }}
\caption{
Net exchange rate between every two levels (x,y), i.e., $X-Y=\dot \rho_{xy}-\dot
\rho_{yx}$, as calculated from the fits of Table $2$. Solid segments are
for positive exchange rate while dashed segments represent redshift
zones where $\dot \rho_{xy} < \dot \rho_{yx}$.
}
\label{fig:net_exchange}
\end{figure}

The evolution of the {\it net~} mass transfer rate between every two components 
($\dot \rho_{xy} - \dot \rho_{yx}$) is shown in Fig.~\ref{fig:net_exchange}. 
Of course, the net transfer can be very different from the actual flows in 
each of the opposite directions. For example, when the flows in the opposite 
directions are equal, the net transfer is zero even when the 
entire population of each component has been exchanged. 
The main lessons from the net transfer rates are as follows:
\begin{enumerate}
\item 
The net exchange U-I at high $z$ is dominated by the accretion from U to I;
it slows down in time while the expulsion from I to U increases.
At $z\simeq0.3$  the backflow becomes comparable to the build-up of I from U.
\item
The net flows from U to S and from I to S are pretty constant at 
$\sim 0.25-0.35$. 
\item 
There is always a significant
net flow from U to D, at the level of $\sim 1$. 
It exceeds the net flow from U to I at $z \leq 1$.
At $z < 0.5$, the evaporation of D to U competes effectively with 
the slowly increasing accretion 
from U to D,
resulting in the decreasing net flow.
\item
The net flow from I to D starts low at early times due to the near absence of 
D material 
at $z>3.5$.
It increases gradually until $z\simeq0.5$ 
by the clustering of I accompanied by tidal stripping. 
At lower $z$ the signal is dominated by artificial fluctuations.
\item
The net exchange between D and S is very low at $z>1$, where
the accretion from D to S is only slightly higher than the tidal stripping
from S to D. At later epochs, the tidal stripping becomes dominant
causing the net exchange to reverse its sign and grow to $\sim 1$ at $z=0$.
\end{enumerate}

In summary, we have presented the mass exchange rates between
several components of cosmological matter in our simulations.
We confirm the generally expected trend that
the mass fraction of ``unvirialized'' 
background material (i.e., all mass outside of any identified halo) 
falls steadily as a function of time, directly
or indirectly fueling the mass accumulation in isolated
halos,  subhalos, and the diffuse media of host halos.  
Although the flow of mass is dominated by the accumulation of
unvirialized background material into and onto isolated halos
and host halos, we find that
there is a  significant amount of mass loss from 
halos back to the unvirialized component and to the diffuse component 
of large host halos.   In the next
section we explore one possible implication of
our derived mass exchange rates.

\section{Integrated History: Metallicity}
\label{sec:metal}

\subsection{Model Predictions}
\label{subsec:metals_mod}

Interesting astrophysical implications may be extracted from
the integrated history of the mass in the different components.
As an example, we describe
an attempt to learn about the possible role of gravitational
effects in determining the metallicity of the gas in the different
components.  We make the crude assumption that the 
baryons trace the
mass distribution everywhere and at all times, and that the 
star formation rate per unit mass and the resulting 
metallicity yield are 
the same in all galactic halos at all times.  
This is based on a scenario in which supernova-driven winds are sufficiently
energetic to drive the gas out of the galactic disks and distribute it in 
quasi-static equilibrium in the galactic halos (e.g., \citeNP{lowferrara}),
but that this feedback is not strong enough
to drive the processed material out of galactic halos 
(e.g., \citeNP{vader:86,ferrara:00}).
The hypothesis we try to evaluate is that
gravitational 
interaction is 
sufficient 
to strip the
enriched matter from galactic halos 
and distribute it in the diffuse components
\cite{david:91,gnedin:98}.

This is of course an extreme hypothesis.  Because star formation rates
grow with the baryon density, the baryons that become stars are highly
concentrated due to gas cooling in the centers of halos; and
supernova-driven winds may not be able to distribute the metals
produced by those stars throughout the halos.  Doubtless in the real
universe there is a mixture of the purely gravitational processes
considered here, on the one hand, and the effects of baryonic physics
(gas cooling, star formation, supernovae), on the other.  In order to
estimate the relative importance of gravitational vs. baryonic
processes, however, it is simplest to consider extreme cases first.

Based on these simplifying assumptions, we virtually assign a fixed 
fraction of gas to each mass particle of the simulation, and associate
with it a metallicity which grows in time in proportion to the time
the particle has spent in a galactic halo.  As defined above,
the galactic halos are
all of the subhalos plus all of the isolated halos.

We focus first on the diffuse component in host halos (D).  We identify 
host halos more massive than $10^{13.6}\hMsun 
= 4\times 10^{13}\hMsun$ with ``clusters",
and those in the range $10^{12}-10^{13.6}\hMsun$ with ``groups".
Although we use the term ``groups'' for this second component,
these objects, especially those in the lower half of the
mass range. are probably better associated with massive galaxies 
hosting satellites than the halos of what are typically referred
to as galaxy groups.  This should
be kept in mind when comparing our predictions to observations,
as discussed in \S4.2.

\begin{figure}
\centerline{\psfig{file=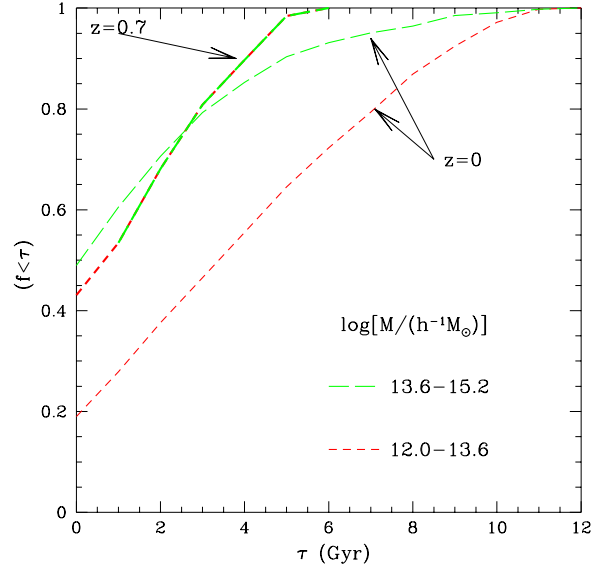
 ,height=8truecm,width=8truecm }}
\caption{
The distribution of time spent by the diffuse components 
in ``galaxies", at $z=0$ (thin) and $z=0.7$ (thick). 
The host halos are divided by mass into ``clusters" and ``groups".
}
\label{fig:time}
\end{figure}

For each host halo, we keep track of the fraction of the 
diffuse mass
that spent less than a given amount of time in galactic halos.
These fractions are averaged over all the host halos of each of the
two mass classes.
Fig. \ref{fig:time} shows the average time distributions at two different 
epochs.  
At $z=0.7$, 
when the $\Lambda$CDM universe considered here had an age of $\sim 7$ Gyr,
the time distributions for the two mass classes are very
similar;
about half of the diffuse component has not spent any time in
galaxies, namely it has been accreted directly from the unvirialized
background. About 5\% has spent more than 4.5 Gyr 
(similar to the solar age) inside galaxies.
At $z=0$, the distributions become quite different.
First, while for the massive hosts the fraction of D material
that has not spent any time in
galaxies is still $\sim 50\%$, this fraction is only $\sim 20\%$ for
the less massive hosts.  Second, while the D faction of massive hosts that
has spent more than 4.5 Gyr inside galaxies is $\sim 10\%$, 
the corresponding fraction of less massive hosts is $\sim 60\%$.
Thus, the diffuse component  in ``clusters" has not been enriched 
significantly by processed material since $z=0.7$, while in ``groups"
it has been enriched significantly during this recent epoch.
This difference hints at a different metallicity enrichment history in 
clusters versus groups.

\begin{figure}
\centerline{\psfig{file=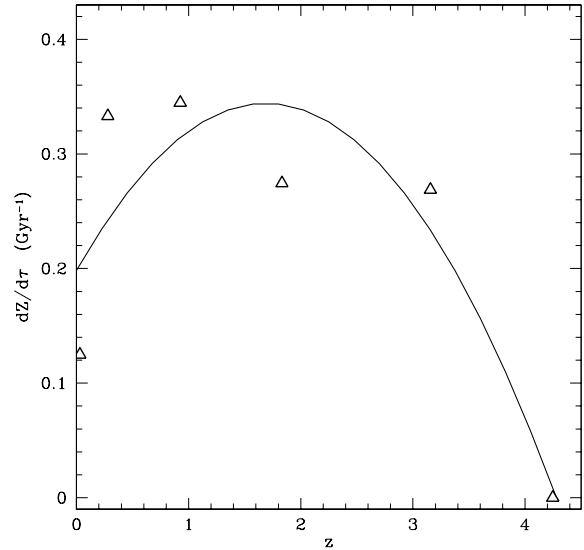
 ,height=8truecm,width=8truecm }}
\caption{
Metallicity production rate versus redshift. The symbols represent time
derivatives of the halo {\it mass weighted} metallicities while the
solid line is a parabolic fit to the equally weighted points (relative
units).
}
\label{fig:metal_prod}
\end{figure}

Figure \ref{fig:metal_prod} shows the overall metallicity production
rate in our scheme as a function of redshift. This is estimated as the
time derivative of the average metallicity, {\it mass weighted} in all halos.
The general behavior of the metal production rate is a rise of about
an order of magnitude between $z\simeq 4$ to $z\simeq 1-1.5$, followed
by a gentler decline of about a factor of $3$ to the rate value at
$z=0$. The interpretation of such a trend in light of observations will
be further discussed in the last section (\S\ref{sec:conc}).

Next, we consider the integrated metallicity in the different components
defined above: the galactic halos I and S, 
the diffuse component D divided by host-halo mass
into ``clusters" and ``groups", and U, the unvirialized IGM.
Since the absolute values of the yield and the gas fraction 
are unknown, we focus on 
the measurement of {\it relative} abundances, between different epochs
or different environments.
Figure \ref{fig:metal} shows the computed evolution of metallicity $Z_x$
in each of these components, in units of the metallicity in clusters at $z=0$,
$Z_{\rm cl}(0)$.

The average metallicity values of the different virialized components grow 
roughly in proportion to time for $z\sim 4$.
In more detail, they all grow between $z=4$ and $2$ at a similar rate
of about $\Delta Z/Z_{\rm cl}(0) = 0.08 \, {\rm Gyr}^{-1}$.
After $z \simeq 2$ (namely, during the last 10 Gyr), the growth rate in
clusters continues at a similar pace, but the growth rate in the
other components speeds up somewhat, to $\Delta Z/Z_{\rm cl}(0) = 0.17,
0.20, 0.24 {\rm Gyr}^{-1}$ for the diffuse component 
in groups, field galaxies (I),
and clustered galaxies (S) respectively.

The average metallicity in the diffuse component in clusters and groups (D) 
is about one half
that in the galaxies (I and S). This is in agreement with our finding in
\S~\ref{sec:rates} that the (enriched) flow from I and S to D is 
roughly comparable to the (fresh) flow from U to D.
 
The metallicity in clustered galaxies (S) is higher than that of field
galaxies (I), by about one third. This is because most S halos are old;
they formed early as I halos, then fell into groups and
clusters, and thus typically had a long time to produce metals.
Many of the current I halos are relatively young, and therefore
less metal rich.
 
The faster growth rate in groups versus clusters at $z<2$
leads to a present metallicity in groups almost twice the
metallicity in clusters.
This is in general agreement with the difference in the time distribution 
seen in Fig.~\ref{fig:time}.

The average metallicity in the unvirialized (U) background is roughly
constant since $z\sim 4$, at a level of 
$Z_{\rm IGM}/Z_{\rm cl}(0) 
\simeq 10^{-4}-10^{-3}$.

\subsection{Comparison with observations}
\label{subsec:metals_obs}

In order to compare our predictions with observations we have to
consider in some more detail the association of the simulated halo 
components with observed objects.
The association of the most massive host halos with galaxy clusters is
natural at all redshifts.
At low redshift, the low mass range of host halos fits well the mass 
range of galaxy groups \cite{hwang:99,davis+mm:99} or 
even massive galaxies with satellite companions. 
It is less obvious what observations to associate with 
the low mass host halos at high redshift. The choice might be
damped Ly$\alpha$ systems (DLAS), or even
Lyman-limit systems and high column density ($\sim
10^{17}$cm$^{-2}$) Lyman-forest clouds.
At high redshift, the unvirialized IGM can be identified with 
Ly$\alpha$ systems of very low column densities, $\sim 10^{14}$cm$^{-2}$, 
which are shown by simulations (\citeNP{dave:98,lu:98}) to correspond to the
mean mass density at that time. 

\begin{figure}
\centerline{\psfig{file=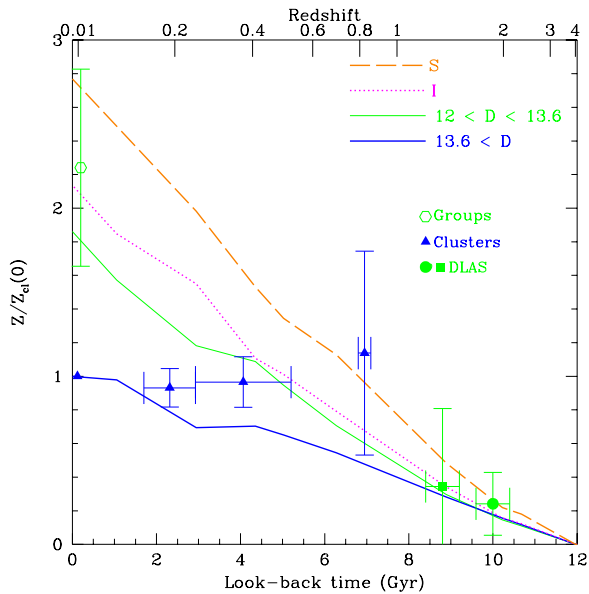
 ,height=8truecm,width=8truecm }}
\caption{
Metallicity of the different components 
as function of look-back time relative to the cluster metallicity
at $z=0$.
Here those halos whose mass lies between $10^{12}$ and $4\times
10^{13} \hMsun$ are called ``groups,'' (thin solid) while those with 
higher mass are called ``clusters'' (thick solid).
The three data points for clusters are from the observed Fe abundance 
in clusters from Mushotzky \& Loewenstein (1997) ($z \simeq 0.2$ and 0.4) 
and Donahue et al. (1999) ($z\simeq0.8$). The data point for groups
(empty hexagon) at $z=0$ (shifted for clarity) is from Buote (2000)
cooling flow model. 
The two data points for DLAS at $z=1-2$ are from Prochaska \& Wolfe (1998)
(circle, higher $z$) and Pettini et al. (1999) (square, lower $z$).
The IGM (U) component metallicity is not shown.
}
\label{fig:metal}
\end{figure}

For clusters today the observational estimate is
$\bar{Z}_{\rm cl}(0) = (0.29 \pm 0.01)Z_\odot$, 
with a relatively small true intrinsic variance between clusters of 
order $0.06 Z_\odot$ \cite{edge+stew:91,yamashita:92}.
We show in Fig.~\ref{fig:metal} three data points with error bars based on
measurements of Fe relative abundance by \citeN{mushotzky:97} 
(at ${\bar z} \simeq 0.2, 0.4$), and \citeN{donahue:00} 
(at ${\bar z} \simeq 0.8$).
Thus, the observed metallicity in the diffuse component of clusters 
does not seem to vary significantly between $z=0$ and 0.4, and not
to change by more than a factor of two out to $z\sim 0.8$.
This is in agreement with the model predictions. 
An even weaker evolution rate is predicted by our model for the most
massive clusters in the simulation, $M>10^{14} \hsolmass$ (not shown in
the figure); this improves the agreement with the observations, for
which massive clusters were preferentially selected,
especially at high redshift.

The observational estimates of the $z=0$ metallicity in groups 
and massive galaxies with satellites
are less secure, and one cannot yet
confirm or refute the predicted trend of 
$Z_{\rm gr}(0) \sim 1.8 Z_{\rm cl}(0) \sim 0.5 Z_\odot$. 
The average metallicity of the diffuse component in rich galaxy groups
is measured locally to be 
$\bar{Z}_{\rm gr}(0) \simeq 0.35 Z_\odot$ \cite{hwang:99}
with the scatter being dominated by an intrinsic variance between
groups of at least $0.1$, partly due to Poisson noise and the small 
number of galaxies per group. 
\citeN{davis+mm:99} 
studied 17 poor groups and found metallicities
of $\sim 0.1 Z_\odot$ for $0.7-1$ keV groups ($\sim 5-8 \times 10^{13} 
\hMsun$) and $\sim 0.2-0.5 Z_\odot$ for more massive groups.  
These values are smaller than our results. 

However recently \citeN{buote:00} showed that for $12$ bright groups,
much higher metallicities are obtained if two-temperature models are
applied with or without cooling flows. The group metallicities increased
from $ {\bar Z}_{\rm gr} = 0.29 Z_\odot$ using a single temperature
model to $0.65-0.75Z_\odot$ depending
on the details of the two temperature models. These values are much 
more consistent with the values presented here.
Perhaps, though, 
a better comparison with our small-mass hosts
is the metallicity estimates of hot gas in massive elliptical galaxies 
at $z=0$.  The measured values \cite{matsumoto:97,buote:98} 
range all the way from 
${\bar Z} = (0.19 \pm 0.12) Z_\odot$ (when single-temperature
models are invoked) through ${\bar Z} = (0.6 \pm 0.5) Z_\odot$ (when
better fits of two-temperature models are considered), and up to 
${\bar Z} = 0.9 Z_\odot$ for the galaxies of best signal-to-noise ratio
\cite{buote:98}. 
The estimate for the metallicity of the Galactic halo interstellar  
gas, however is lower \cite{savage:96} and stands on about $0.1$ 
solar for most elements. 
The observed metallicities 
would be consistent with our predictions for 
${\bar Z}_{\rm gr}(0)/{\bar Z}_{\rm cl}(0)$ if a substantial fraction of
the less massive host halos which we classified as ``groups" 
actually correspond to the massive galaxies of $Z \sim 0.6-0.9Z_\odot$.

If DLAS are the counterparts of our ``group'' halos  at high redshift, then
the relevant metallicity measurements \cite{lu,pw:98} give 
${\bar Z}_{\rm DLA} \simeq 0.07 Z_\odot$ at $z\ga 1.5$. 
The measurement for a sample of DLAS with $z\la 1.5$
\cite{pettini:99} is ${\bar Z}\simeq 0.1 Z_\odot$.
These data points are shown in Fig.~\ref{fig:metal}.
When compared to today's cluster metallicity,
the observed values correspond to 
$\bar{Z}_{\rm gr}(z \sim 1.5)/\bar{Z}_{\rm cl}(0) \sim 0.2$, 
in surprisingly good agreement with the model predictions.
If, alternatively, we associate DLAS at high $z$ with 
groups and very massive galaxies with satellites 
at $z=0$ (both have similar hydrogen column densities in the range
$10^{20}-10^{21}$ atoms cm$^{-2}$), we obtain a ratio
$\bar{Z}_{\rm gr}(z \sim 1.5)/\bar{Z}_{\rm gr}(0) \sim 0.26$,
also in agreement with the model predictions.
However, there are certain caveats associated with this comparison.
For example, unlike the measurements in groups at $z=0$,
the observed metallicities at high $z$ were not evaluated by Fe abundance.
Also, these abundances refer in large to the cold gas component,
and are relevant to our model predictions only if the assumption 
of proper mixing between the cold disk and the hot halo is valid.
Finally, the large scatter in these measurements at high $z$
weakens any conclusion drawn on the basis of the mean values.

The predicted metallicity of the unvirialized IGM may be compared
to the very low column density Ly$\alpha$ clouds. Data for abundances of
these exist only for $z\simeq2.2-4$ \cite{savaglio:97,dave:98,lu:98}. 
These results show metallicities in the range 
$0.3-3 \times 10^{-4}$.
Since no significant systematic evolution in metallicity is observed in this
redshift range, these values may be indicative of lower redshift
values as well. The simulation analysis shows little evolution 
in the metallicity of the U component, and fluctuations in the range 
$Z_{\rm IGM} = 0.3 - 1 \times 10^{-3} Z_{\rm cl}(0) \sim 1-3
\times 10^{-4} Z_\odot$. This agrees
well with the observed values.

\section{Conclusion and Discussion}
\label{sec:conc}

We have learned that
the role of gravity does not end when a virialized halo is formed.
Gravity is responsible for a continuous and substantial exchange of 
matter among the different components of halos and diffuse media.
This exchange may affect the formation and evolution of luminous
galaxies inside the halos and their feedback into the environment. 
It should therefore be included in the modeling of structure formation.
We have analyzed this exchange between 
4 basic components of
DM halos and diffuse media, and provided 
quantitative exchange rates
to be used in further investigations.
We then showed, in particular, that gravitational 
interaction and stripping 
may provide the mechanism which produces 
a significant part of 
the relatively high 
metallicities observed in the diffuse hot 
gas of big galaxies, groups and clusters 
as well as the 
unvirialized background IGM. 

A summary of the matter division and exchange rates is as follows.
Most 
of the mass within identified halos ($\ga 10^{11} \hMsun$) is
located in ``isolated'' galactic halos 
before $z \sim 2$.  Then about 15\% of 
their mass 
accumulates 
in groups and clusters 
per Gyr, while a similar fraction is expelled back to the 
unvirialized background.  
Subhalos tend to form later, mostly by a constant-rate 
accretion from outside the clusters, but at $z \leq 1$ 
they also accrete 
from the cluster diffuse medium. The subhalos are constantly
disrupted such that about one half the subhalo mass is being exchanged 
every Gyr.  The diffuse medium in clusters is built by accreting 
extra-cluster material, composed on average of 40\% halos and 60\% 
unvirialized matter. The disruption of subhalos becomes a significant
source after $z \sim 0.5$.
Finally, the unvirialized IGM is enriched mostly by mass loss from
isolated galactic  
halos, and at late epochs, $z<1$, also by expulsion from clusters.

The results summarized above highlight important limitations of the recipes
used by current semi-analytic models \cite{kwg:93,cafnz:94,sp:99}, 
and offer possible routes
for improvement.  In particular,
\begin{itemize}
\item
Matter expulsion from isolated halos, which is found here to be quite 
significant, is completely ignored in the semi-analytic models because they are based on 
a hierarchical clustering formalism which does not 
take into account such 
a process.  Tidal stripping and gravitational heating in collisions
are known to be more effective in denser environments and at
higher redshifts \cite{collision_rate}.
While the time dependence can be incorporated in semi-analytic models quite straightforwardly,
the environment dependence is harder to mimic.
\item
We notice that ``fresh" matter from the unvirialized component (U) 
finds its way to subhalos (S),
while current semi-analytic models stop the supply of DM and 
fresh gas to halos as soon as they 
are incorporated in bigger halos, even when the subhalo and the host
are of comparable masses.
\item
We find that accretion from diffuse matter in halos (D) 
onto subhalos (S) becomes the main source of matter for S
at low redshift, while 
this process is completely ignored in semi-analytic models.
Some semi-analytic models do allow for tidal stripping of subhalos, but
the matter exchange between subhalos and their hosts 
seems to be 
a two-way process that should be treated as such in the semi-analytic models.
\item 
The fitting formulae 
Eq. (2) and Table 1
describing the exchange rates as a function of time
can be used as a first attempt to improve semi-analytic treatment of matter exchange.
Since we only provide average exchange rates without specifying their
dependence on halo mass, a sensible utilization of these formulae is
via the implied fractional exchange rates rather than the absolute rates.
\end{itemize} 

A convenient use of these formulae may be a direct integration to yield
the density of each component at an arbitrary time. Integrations
of the rate equations in the redshift range of $[4-0]$ yield results
accurate to about $20\%$ in comparison to the actual densities at the
upper limit of the integration time. 
The initial conditions were taken to be the actual density
of the relevant component at the lower limit of the integral (higher
redshift). Both the upper and lower limits of the integrals can be read
directly from Table $1$.

The second part of this paper is a more speculative attempt to
demonstrate how the matter exchange may affect measurable quantities.
This is an
attempt to predict metallicities from simulations
that do not include explicitly any gas, by simply assuming the metallicity 
of the different components is proportional to the time this matter spent 
in galactic halos.
We effectively assume
that supernova
winds expel
processed matter 
from the disk out 
into each halo, so that 
this processed matter spreads
uniformly in it. However we assume that
the SN winds are not powerful enough to drive
the matter all the way out of individual halos. 
We address whether gravitational interactions could have
acted to 
provide the missing energy and thus to 
enrich the IGM with this material.
We avoided any calibration of the specific star formation rate and 
yield in absolute
terms, and instead addressed relative metallicities, comparing different 
times or different environments.

We found that the relative metallicities, compared to that of
galaxies, 
obtained by gravitational 
processes in the diffuse components of groups, clusters, and the IGM
are in the same ball park as the observed 
metallicities at $z=0$.
This scenario seems to predict an anti-correlation between the metallicity
and the mass of the host halo, which is yet to be tested against
observations. For example, it may be possible to measure halo masses
at high z 
from line profiles \cite{pw:97a,pw:98a,haehnelt:97,maller:00} and thus allow
a direct confrontation of prediction and observation.
The predicted slow evolution of metallicity in clusters from
$z=1$ to 0 is consistent with observations.
Matching the simulated halos with real objects at high redshifts
is more ambiguous; if DLAS or Ly$\alpha$ systems are 
the high-redshift counterparts of local groups and very massive 
galaxies with satellites, then the predictions agree with the observed
metallicity evolution.
Our tentative conclusion is that gravitational effects
could indeed be largely responsible for bringing  
the metals from galactic halos into the diffuse media.

The process of 
a gravitational merger mechanism for gravitational enrichment
has already been addressed by
Gnedin (1998), in the context of interacting field galaxies and the IGM,
using $\leq3 \hmpc$ simulations in which sub-cell physics was added ``by hand".
Our N-body results in a cosmological volume confirm the feasibility of
this general idea and demonstrate that gravity may be responsible for 
the redistribution of metals in the diffuse component of clusters and 
groups as well.
The very low metallicity predicted by our analysis for the unvirialized
IGM compared to the virialized components is in agreement with the 
predictions based on the 
abovementioned 
simulations.

A more elaborate model for the hot gas and metal redistribution in
clusters, groups, and massive ellipticals
\cite{mathews:98,brighenti:99b,brighenti:99a} have demonstrated some of the
difficulties in combining a mixture of supernovae as the 
source of metals and enrichment of the diffuse media. 
While these models succeed in reproducing the observed abundances of hot
gas in elliptical galaxies, the straight sum of the predicted enrichment 
from the member galaxies fails to account for the observed metallicity 
of the diffuse media in groups and clusters.
However, close encounters of galactic halos and the associated
harassment
(\citeNP{moore:99})
are more frequent in the crowded environment of groups and clusters 
\cite{collision_rate} and may thus provide the missing ingredient 
in these models.

While we mainly addressed the average metallicity in the different
environments, the scatter about the mean may also be of interest.
A large scatter in metallicity at a given density has been stressed
by \citeN{cen+ost:99}, but the mass bins used here can be more 
easily identified with real objects,
and thus allow a more direct comparison with observations. 

Of course, our attempt to estimate metallicities is extremely simplified
and it ignores many complexities involving gas dynamics and feedback
that should be properly addressed in more detailed studies.

For example, the gas cannot be uniform and cannot trace perfectly the 
dark-matter distribution. 
The cooling time is in fact different at
different positions in the halo and at different times, and it 
is likely to develop the two phases of cold clouds in a hot medium.
The feedback may not be strong and persistent enough to keep the
expelled gas hot and transported to large radii in the galactic halos,
where most of the stripping is expected to occur.

In reality, the metallicity yield must vary in time and in space.
A specific worry is associated with the fact that the star-formation
rate (SFR) is known to drop at late epochs, $z<1$ 
\cite{madau:97,spf:00}. 
This is largely a result of the gas processes, such as cooling, that
determine the characteristic upper limit of galactic masses,
which we do not explicitly simulate.
Furthermore, we do not explicitly take into account the effect of
gas depletion in old halos on the SFR.
Two features of our analysis ease this problem.
First, 
our analysis does mimic a moderate drop in the SFR at late times,
if SFR is taken to be the net flow rate (time derivative of the density)
to each galactic component. This drop is due to the
combination of three effects: 
(a) the slow down in the growth of fluctuations due to the \lcdm\
cosmology, (b) the imposed minimum on the halo masses, and (c) the 
identification of only simple halos with galactic halos.
Second, the yield of metals is expected to vary slower than the SFR
because it results from stars of different ages.

A related, interesting comparison to observations can be made by the
examination of the overall metallicity production rate in view of the
SFR diagrams as obtained from observations (the ``Madau diagrams''). 
A direct comparison between the metallicity production rate in our
recipe (Fig. \ref{fig:metal_prod}) and recent versions of SFR diagrams
(see Somerville et al. 2000 for references)
indicate two distinct differences: (i) The SFR peak is
obtained at a higher redshift ($z\simeq2.5-3$) than in our recipe for
the metal production ($z\simeq 1-2$) (ii) the drop from the SFR peak 
value to its $z=0$ value is slightly bigger ($\sim 3-4$) than our
value ($ \sim 2-3$).
However the link between an SFR diagram and metallicity production
diagrams involves the assumption that all metals are produced by
type II supernovae which closely trace the SFR. If in contrast, a large
fraction of metals (especially Iron, from which the cluster metallicity
is calculated) is produced in type Ia supernovae then the one to one
mapping between SFR and metallicity production becomes less clear.
Such SNIa contribution to the metal production is now favored by
observations (cf. \citeNP{renzini:00}). In a scenario where combined
contributions from different SN types dictate the metallicity production
rate, the observational SFR diagrams can be easily reconciled with the
metallicity production rate as derived by the proposed recipe of this paper.

Our comparison with metallicity observations is also very simplified. 
For example, our simple model provides the volume-average metallicity 
within the host halos, but in reality one may expect spatial gradients 
in metallicity.  The observations, which are sometimes directed to 
different positions in the halos, e.g., to cluster centers,
may therefore yield different metallicities.
Another complication is that the metals may be shared between (at least) 
two phases of gas, cold and hot, as well as the stellar and dust components.

Our crude treatment of one single ``metallicity" is an over-simplification, 
because actual metallicity measurements 
refer to a variety of elements,
usually different elements at different redshifts 
depending on the detectability of the corresponding spectral lines. 
Moreover, since different elements in different abundances are produced
through a 
variety of mechanisms \cite{mathews:98,brighenti:99b,brighenti:99a},
the detailed history of various processes may be relevant for the metal
production. A similar difficulty arises due to the (sometimes unknown)
ionization state of the medium in which metallicities are observed,
because in our simple model we cannot determine 
the gas temperature.

Our metallicity estimates should therefore
be considered very crude
and serve only as an encouraging indication for the possible important role
of gravity in the large-scale exchange of metals.

It will be very interesting to see how important ``nature'' (dark
matter physics) is compared to ``nurture'' (baryonic physics).  The
fact that the Madau-like plot (Figure \ref{fig:metal_prod}), 
based on including only
gravitational effects, looks reasonably similar to the observationally
based plot is encouraging.  Semi-analytic modelers should accept the
challenge to disentangle the relative roles of ``nature'' vs. ``nurture.''

\section*{Acknowledgments}

The  simulations  were  performed  at NRL  and   NCSA.  This  work was
supported  by  the  US-Israel Binational   Science  Foundation, by the
Israel Science Foundation, and by grants from NASA and NSF at UCSC and
NMSU.  J.S.B. was supported in part by NASA LTSA grant 
NAG5-3525 and NSF grant AST-9802568.  Support for 
A.V.K.  was  provided by NASA through Hubble
Fellowship  grant HF-01121.01-99A  from   the Space Telescope  Science
Institute, which is operated   by the Association of  Universities for
Research in Astronomy, Inc., under NASA contract NAS5-26555.

\bibliographystyle{mnras}
\bibliography{mnrasmnemonic,all_refs}
\end{document}